\documentclass[12pt]{iopart}
\pdfoutput=1
\usepackage{color}
\usepackage{amssymb}
\usepackage{graphicx}

\def\beqra{\begin{eqnarray}}
\def\eeqra{\end{eqnarray}}
\def\beq{\begin{equation}}
\def\eeq{\end{equation}}

\def\etain{\eta_{in}}

\def\vp{\bar{\varphi}}

\def\vp{\varphi}

\def\bx{{\bf{x}}}
\def\bk{{\bf{k}}}
\def\bp{{\bf{p}}}
\def\bq{{\bf{q}}}

\def\bV0{{\bf{V_0}}}
\def\re#1{(\ref{#1})}

\def\agt{~\mbox{\raisebox{-.6ex}{$\stackrel{>}{\sim}$}}~}
\def\alt{~\mbox{\raisebox{-.6ex}{$\stackrel{<}{\sim}$}}~}
\def\bx{{\bf{x}}}

\def\bk{{\bf{k}}}
\def\bp{{\bf{p}}}
\def\bq{{\bf{q}}}

\def\vpb{\bar{\vp}}
\def\vp{\varphi}

\begin{document}
\begin{flushright}
{\small UMN-TH-3628/17}
\end{flushright}

\title{A Robust BAO Extractor }

\author{Eugenio Noda$^{1,2}$, Marco Peloso$^{3}$, Massimo Pietroni$^{1,4}$}
\vskip 0.3 cm
\address{
$^1$Dipartimento di Scienze Matematiche, Fisiche ed Informatiche dell'Universit\`a di Parma, Parco Area delle Scienze 7/a, I-43124 Parma, Italy\\
$^2$INFN, Gruppo Collegato di Parma, Parco Area delle Scienze 7/a, I-43124 Parma, Italy\\
$^3$School of Physics and Astronomy, and Minnesota Institute for Astrophysics, University of Minnesota, Minneapolis, 55455, USA\\
$^4$INFN, Sezione di Padova, via Marzolo 8, I-35131, Padova, Italy\\
}

\begin{abstract}
We define a procedure to extract the oscillating part of a given nonlinear Power Spectrum, and derive an equation describing its evolution including the leading effects at all scales. The intermediate scales are taken into account by standard perturbation theory, the long range (IR) displacements are included by using consistency relations, and the effect of small (UV) scales is included via effective coefficients computed in simulations. We show that the UV effects are irrelevant in the evolution of the oscillating part, while they play a crucial role in reproducing the smooth component. Our ``extractor'' operator can be applied to simulations and real data in order to extract the Baryonic Acoustic Oscillations (BAO) without any fitting function and nuisance parameter. We conclude that the nonlinear evolution of BAO can be accurately reproduced at all scales down to $z=0$ by our fast analytical method, without any need of extra parameters fitted from simulations.
\end{abstract}

\maketitle

\section{Introduction}
The evolution of the large scale structure of the Universe (LSS) beyond the linear regime must be accurately described in order to test cosmological models with present and forthcoming observations. Besides N-body simulations, analytical tools have been investigated and developed in the last years (for a review, see \cite{Bernardeau:2013oda}). It is well known that standard perturbation theory (SPT) \cite{PT} fails to achieve  percent accuracy at low redshifts and at BAO scales, and smaller. The failure is due to the inaccurate description of the couplings of the intermediate scales with very large (IR) and very small (UV) ones. 

At large scales, the dominant effect is given by long range displacements, of typical rms $\sim 6 ~\mathrm{Mpc/h}$, induced by velocity fields  coherent on scales as large as $\sim110~ \mathrm{Mpc/h}$. This effect can be well described in Lagrangian Perturbation Theory, already at the level of the Zel'dovich approximation, but is badly reproduced by lowest order SPT. Methods to overcome this problem have been envisaged, and after some debate regarding the Galilean invariance of the proposed schemes, a consensus has emerged on the way to resum the leading IR effects at all SPT orders \cite{Senatore:2014via,Baldauf:2015xfa,Blas:2016sfa,Peloso:2016qdr}. 

The effect of very small scales on intermediate ones can be studied by looking at the ``response function" , which quantifies the sensitivity of the nonlinear Power Spectrum (PS) at a given scale $k$ on a slight modification of the linear one at a different scale $q$. N-body simulations show a progressive screening of the effect of UV scales on intermediate ones, a behavior which is completely missed by SPT \cite{Nishimichi:2014rra}. The UV failure of SPT is expected, as nonlinear effects become more and more relevant at small scales and, at a more fundamental level, it is based on the single-stream approximation, in which velocity dispersion and all higher order momenta of the Dark Matter (DM) particle distribution function are artificially set to zero. 

The shortcomings of SPT in describing the effect of small scales can be overcome by methods such as the Coarse Grained Perturbation Theory (CGPT)  \cite{Pietroni:2011iz,Manzotti:2014loa} or the Effective Field Theory of the Large Scale Structure EFToLSS \cite{Baumann:2010tm, Carrasco:2012cv} (see also \cite{Floerchinger:2016hja}), in which the ``wrong'' UV behavior of SPT is replaced by counterterms measured from N-body simulations. These procedures have been shown to increase the maximum wavenumber $k_{max}$ up to which they match the results from simulations from $k_{max}^{SPT}\sim 0.05 \;\mathrm{h/Mpc}$ to, typically, $k_{max}\sim 0.25-0.3 \;\mathrm{h/Mpc}$ at $z=0$. Moreover, in \cite{Manzotti:2014loa} it was shown that the cosmology dependence of these counterterms can be efficiently reproduced by SPT, so that one does not need to run a simulation for each different cosmological model. 

In this paper we present a scheme in which the three physical effects discussed above are clearly identified and combined. The mode-coupling between nearby intermediate scales is taken into account by SPT, the effect of UV scales is treated via effective counterterms and the long range displacements are taken into account by IR resummations based on nonlinear consistency relations. Our scheme improves the Time Renormalization Group (TRG) of ref.~\cite{Pietroni08} and the time-evolution equations of refs.~\cite{Anselmi:2010fs,Anselmi:2012cn} by consistently including these effects while conserving the flexibility of these approaches in describing, for instance, models in which the scale factor is scale dependent, as in massive neutrino cosmologies \cite{LMPR09}.

By clearly identifying the different physical effects we are able to assess the dependence of a given observable on each of them. Therefore, we focus on BAO, and derive an evolution equation for the oscillating component of the PS alone. By using exact consistency relations first derived in \cite{Peloso:2013zw,Kehagias:2013yd}, we include the effect of long range displacements at all SPT orders. Moreover, the structure of the equation itself suggest a procedure to ``extract'' the oscillating part from a given nonlinear PS, as the part of the PS which evolves independently from the smooth one. This 
``extractor operator'' can be also applied to the PS obtained in N-body simulations and in real data, to directly compare theory and observations. This is to be contrasted with the usual approach, in which to extract the BAO scale one usually fits the data with a function containing a number of nuisance parameters modelling the nonlinear evolution of the broadband spectrum and the damping of the BAO fluctuations. 

Comparing our extracted BAO with simulations we can clearly show that the BAO evolution is largely insensitive to the details of UV physics: our IR resummed equation for the oscillating part accurately describes BAO at all scales and down to $z=0$, and the results do not change appreciably even by setting to zero the UV counterterms (in EFToLSS language, the ``sound speed'' and other coefficients). The latter are, on the other hand, crucial in order to reproduce the broadband spectrum, and we detail the procedure to be followed to systematically include them at higher orders.

The paper is organised as follows: in sect.~\ref{eveq} we introduce the evolution equations derived in refs.~\cite{Pietroni:2011iz,Manzotti:2014loa} and introduce the formalism and notations, in sect.~\ref{UVsect} we discuss the inclusion of UV effects and show the results for the broadband PS at the lowest nontrivial order in our approximations. In sect.~\ref{IRsum} we focus on the BAO and derive an evolution equation for the oscillating part of the PS including large scale flows by means of consistency relations. In sect.~\ref{BAOext} we define our ``BAO extractor'' and apply it both to our results and to the PS from N-body simulations, discussing the comparison between the two. Finally, in sect.~\ref{conclusions} we summarise our results and present our outlook on future work. In \ref{app:eta-C} we describe the procedure for extracting the UV counterterms.

\section{Evolution equation for the PS}
\label{eveq}
Neglecting vorticity, the first two moments of the coarse-grained Vlasov equations \cite{ Pietroni:2011iz, Manzotti:2014loa} give the continuity and Euler equations, which, in Fourier space, can be cast in compact form as 
\beqra
&&\!\!\!\!\!\!\!\!\!\!\! \!\!\!\!\!\!\!\! (\delta_{ab}\partial_\eta +\Omega_{ab})\vp_b^R(\bk,\eta)= I_{\bk,\bq_1,\bq_2}  e^\eta\gamma_{abc}(\bq_1,\bq_2) \vp_b^R(\bq_1,\eta)\vp_c^R(\bq_2,\eta) -h_a^R(\bk,\eta) \,,
\label{eom}
\eeqra
where $\eta=\log D(\tau)$ and $I_{k;p_1,\cdots,p_n} \equiv \int \frac{d^3p_1}{(2 \pi)^3} \cdots \frac{d^3p_n}{(2 \pi)^3} (2\pi)^3 \delta_D(\bk-\sum_{i=1}^n \bp_i)$.

We have introduced the fields
\begin{equation}
\!\!\!\!\!\!  { \vp}_1^R \left( \bk ,\eta \right) \equiv {\rm e}^{-\eta} {\bar \delta_R } \left(  \bk,\eta \right) \;\;,\qquad\quad
{\vp}_2^R \left( \bk,\eta \right) \equiv  {\rm e}^{-\eta}  \frac{-\bar \theta ^R \left(  \bk,\eta \right) }{{\cal H } f}\,,
\label{phi1-2}
\end{equation}
where we have also used the linear growth function $f \left( \eta \right) \equiv \frac{1}{\cal H} \frac{d \eta}{d \tau}$.  Repeated indices are summed over $b,c=1,2$. The overbars indicate  fields obtained by  Fourier transforming $[ \delta ]_R (\bx,\eta)$ and the divergence of $[(1+\delta) v^i ]_R (\bx,\eta)/[ 1+\delta  ]_R (\bx,\eta)$, respectively, where $[ \cdots ]_R$ indicates filtering up to some scale   $R$ (see sect.~\ref{UVsect} and refs.~\cite{ Pietroni:2011iz, Manzotti:2014loa} for details).
The left hand side of this relation is the linearized part of the equation for the two dynamical modes, and it is characterized by the matrix
\begin{equation}
{\bf \Omega} = \left( \begin{array}{cc} 1 & - 1 \\ - \frac{3}{2} \frac{\Omega_m}{f^2} &  \frac{3}{2} \frac{\Omega_m}{f^2} \end{array} \right) .
\label{bigomega}
\end{equation}
The first term at the right hand side of eq. (\ref{eom}) encodes the mode-coupling between the  fields. The only nonvanishing components of the vertex functions are 
\begin{equation}
\!\!\!\!\!\! \!\!\!\!\!\! \!\!\!\!\!\! \gamma_{121} \left( \bp , \bq \right) = \frac{\left( \bp+\bq \right) \cdot \bp}{2 p^2} \;\;,\;\;
\gamma_{112} \left( \bq , \bp \right) = \gamma_{121} \left( \bp , \bq \right) \;\;,\;\; 
\gamma_{222} = \frac{\left( \bp + \bq \right)^2 \bp \cdot \bq}{2 p^2 q^2} \,.
\label{gamma}
\end{equation} 
The mode-coupling term can also be written as
\beqra
&&\!\!\!\!\!\! \!\!\!\!\!\! \!\!\!\!\!\! e^\eta\gamma_{abc}(\bq_1,\bq_2)  \vp_b^R(\bq_1,\eta)\vp_c^R(\bq_2,\eta)= e^\eta \delta_{a2}\,\tilde\gamma (\bq_1,\bq_2) \vp_2^R(\bq_1,\eta)\vp_2^R(\bq_2,\eta)\nonumber\\
&& \qquad\qquad\qquad\qquad\qquad-i k^j\frac{\bar v^j_R(\bq_1)}{{\cal H}f} \vp_a^R(\bq_2,\eta)\,,
\label{IRsplit}
\eeqra
with
\beq
\tilde\gamma (\bq_1,\bq_2) =\frac{(\bq_1\cdot\bq_2)^2}{q_1^2q_2^2}-1\,. 
\label{resvert}
\eeq
When  $q_1\ll k\simeq q_2$ the second term at the RHS of \re{IRsplit} singles out the leading contributions to the mode-coupling induced by long (and time-dependent) velocity modes, whose form is dictated by the Galilean invariance of the system \cite{Scoccimarro:1995if,Peloso:2013zw,Peloso:2013spa}. Expressing the velocity through its divergence, it gives the (formally) IR divergent term
\beq
e^\eta \frac{\bk\cdot\bq_1}{q_1^2}\vp^R_2(\bq_1,\eta)\vp_a^R(\bq_2,\eta)\,,
\label{IReff}
\eeq
where we have used the definition \re{phi1-2}.
In sect.~\ref{IRsum} we will discuss the resummation of these effects at all orders. On the other hand, the residual vertex function, eq.~\re{resvert}, amounts to $\cos(\theta_{12})^2-1$, where $\theta_{12}$ is the angle between $\bq_1$ and $\bq_2$, and therefore it is never divergent, and moreover vanishes for $\bk=\bq_1+\bq_2\to 0$.

The last term on the right hand side of  eq. (\ref{eom}) is the contribution from the short modes that have been integrated out in the coarse-graining procedure \cite{ Pietroni:2011iz, Manzotti:2014loa},
\beq
h_a^R(\bk,\eta) \equiv -i\,\frac{k^i J^i_R(\bk,\eta)}{{\cal H}^2 f^2}e^{-\eta} \delta_{a2}\,,
\label{htot}
\eeq
where the UV source $J_R^i(\bk,\eta)$ is the Fourier transformed of 
\beq
J^i_R(\bx,\eta)= J_{1,R}^i(\bx,\eta)+J_{\sigma,R}^i(\bx,\eta)\,,
\eeq
which depend on the gravitational potential $\phi$  and the velocity dispersion $\sigma_R^{ik}$,
\beqra
&&
J_{1,R}^i(\bx,\eta)=\frac{1}{[ 1+\delta]_R(\bx,\eta)} [ ( 1+\delta) \nabla^i\phi ]_R(\bx,\eta) - \nabla^i [ \phi]_R (\bx,\eta)\,,\nonumber\\
&&J_{\sigma,R}^i(\bx,\eta)=\frac{1}{[ 1+\delta]_R(\bx,\eta)}\frac{\partial }{\partial x^k} \left[  [1+\delta]_R(\bx,\eta) \sigma_R^{ik}(\bx,\eta)\right]\,.
\label{UVsources}
\eeqra

Applying the equation of motion \re{eom} to the (equal-time) PS,
\beq
P_{ab}^R(k) = \langle \vp^R_a(\bk) \vp^R_b(-\bk) \rangle'\,, 
\eeq
-- where the prime indicates that we have divided by $(2\pi)^3$ times the overall momentum delta function --
 gives
\beqra
&&\!\!\!\!\!\! \!\!\!\!\!\! \!\!\!\!\!\! \!\!\!\!\!\! \!\!\!\!\!\! \!\!\!\!\!\! \partial_\eta  P^R_{ab}(k) = \Bigg[ -\Omega_{ac}  P^R_{cb}(k)  + e^\eta I_{\bk;\bp_1,\bp_2} \gamma_{acd}(\bp_1,\bp_2)B^R_{bcd}(k,p_1,p_2) - \langle h^R_a(\bk) \vp^R_b(-\bk) \rangle' \nonumber\\
&&+(a\leftrightarrow b)\Bigg],
\label{TRG1}
\eeqra
where we have omitted the $\eta$-dependence, and where the bispectrum is given by
\beq
B^R_{abc}(q_1,q_2,q_3) =  \langle \vp^R_a(\bq_1) \vp^R_b(\bq_2) \vp^R_c(\bq_3) \rangle'\,.
\eeq

Before proceeding, we emphasize that the only approximation in the equation above is in the way we deal with the vorticity of the coarse-grained velocity field. This has two components: a microscopic one, related to UV scales smaller than $R$, and one induced by the coarse-graining procedure itself. While the first one is completely included in the source terms $h_a^R$, we deal with the second one at a perturbative level.  While in this section we have set the second vorticity component to zero from the beginning, one can show, using the methods of \cite{Manzotti:2014loa}, that including it perturbatively would give exactly the same equations as those considered in the next sections.
The effect of vorticity on the PS was investigated in \cite{Pueblas:2008uv}, where it was found to be negligible at all scales and redshifts of interest.

No other approximation has been imposed so far. In particular, we are not assuming the single stream approximation, as it is usually done in SPT and other semi-analytic methods. Eq.~\re{eom}, and its PS counterpart, eq.~\re{TRG1}, contain all the relevant physics: the effect of the UV scales on the intermediate ones, through the source $h_a^R$, the mode-coupling between the intermediate scales, through the vertex functions, and the IR displacements in the terms containing the vertex \re{IReff} for $q_1\ll k$. In the following, we will discuss how to deal with all these effects.

\section{UV effects}
\label{UVsect}
The source term $h_a^R$ is responsible for all deviations from the single stream approximation and all the nonlinear effects occurring at small scales. It is therefore cleaner to consider the subtracted PS,
\beq
\Delta P^R_{ab}(k) \equiv P^R_{ab}(k)- P^{R,ss}_{ab}(k)\,,
\eeq
where $P^{R,ss}_{ab}(k)$ is the PS computed in the single stream approximation. It solves an equation analogous to \re{TRG1}, in which all the quantities, including the $\langle h_a^R \vp_b^R\rangle^\prime$ correlator are obtained in the single stream approximation, in practice, by considering SPT or other approximation schemes at some finite order. This correlator vanishes in the $R\to 0$ limit while its value at nonvanishing $R$ takes into account all nonlinear effects due to modes $q\agt 1/R$ in the single stream approximation, to be subtracted from the fully nonperturbative correlator measured in N-body simulations. Working at finite $R$ is essential in practice, in order to extract the source terms from the simulation, but the final results should be independent on the value chosen for $R$. We discuss this point below.

 The evolution equation for the subtracted PS therefore reads 
\beqra
&&\!\!\!\!\!\! \!\!\!\!\!\! \!\!\!\!\!\! \!\!\!\!\!\! \partial_\eta \Delta P^R_{ab}(k) = \Bigg[ -\Omega_{ac}\Delta P^R_{cb}(k)  + e^\eta I_{\bk;\bp_1,\bp_2} \gamma_{acd}(k,p_1,p_2)\Delta B^R_{bcd}(k,p_1,p_2)\nonumber\\
&& \qquad - \Delta \langle h_a^R(\bk) \vp^R_b(-\bk) \rangle' +(a\leftrightarrow b)\Bigg],
\label{evDP}
\eeqra
where
$ \Delta \langle h_a^R(\bk) \vp^R_b(-\bk) \rangle'  \equiv  \langle h_a^R(\bk) \vp^R_b(-\bk) \rangle' - \langle h_a^{R,ss}(\bk) \vp^{R,ss}_b(-\bk) \rangle' $. The subtracted bispectrum,
$\Delta B^R_{abc}(k,p_1,p_2)\equiv  B^R_{abc}(k,p_1,p_2)- B^{R,ss}_{abc}(k,p_1,p_2)$,
in turn, solves the equation
\beqra
&&\!\!\!\!\!\! \!\!\!\!\!\! \!\!\!\!\!\! \!\!\!\!\!\! \partial_\eta \Delta  B^R_{abc}(k,q,p) = \nonumber\\
&&\!\!\!\!\!\! \!\!\!\!\!\! \!\!\!\!\!\!  \Bigg[- \Omega_{ad}\Delta B^R_{dbc}(k,q,p) - \Delta\langle  h_a^R(\bk) \vp^R_b(\bq) \vp^R_c(\bp) \rangle' \nonumber\\
&&\!\!\!\!\!\! \!\!\!\!\!\! \!\!\!\!\!\! +2 e^\eta \gamma_{aef}(k,q,p) \left(\Delta P^R_{eb}(q)  P^{R,ss}_{fc}(p)  +  P^{R,ss}_{eb}(q)\Delta  P^{R,ss}_{fc}(p)  + \Delta  P^R_{eb}(q)\Delta  P^R_{fc}(p) \right)
\nonumber\\
&&\!\!\!\!\!\! \!\!\!\!\!\! \!\!\!\!\!\! + e^\eta I_{\bk;\bp_1,\bp_2} \gamma_{ade}(k,p_1,p_2) \Delta T^R_{debc}(\bp_1,\bp_2,\bq,\bp) \nonumber\\
&&\!\!\!\!\!\! \!\!\!\!\!\! \!\!\!\!\!\! + {\mathrm{cyclic \;\;permutations\;\;of}}\;\; (a,\bk),  \; (b,\bq),\; (c,\bp)\Bigg]\,,
\label{DB}
\eeqra
where $ \Delta\langle  h_a^R \vp^R_b \vp^R_c \rangle'\equiv \langle  h_a^R \vp^R_b \vp^R_c \rangle'-\langle  h_a^{R,ss} \vp^{R,ss}_b \vp^{R,ss}_c \rangle'$ and we have defined the deviation from the single stream approximation of the trispectrum, $ \Delta T^R_{debc}\equiv   T^R_{debc}-  T^{R,ss}_{debc}$, which, in turn, solves an evolution equation which can be straightforwardly derived. 

The system of  coupled evolution equations solved by the subtracted correlation functions must be truncated at some order. However, unlike the original TRG proposal \cite{Pietroni08}, where the evolution equations for the unsubtracted functions were considered, there is a clear hierarchy in these equations which leads to a natural criterium for the truncation. Indeed, we first notice that the sources of these equations are given by the $ \Delta\langle  h_a^R \vp^R_{b_1} \vp^R_{b_2}\cdots \vp^R_{b_n} \rangle'$ correlators, since, if they all vanish,  the single stream approximation is exact. Moreover, 
the role of these differences between correlators is to replace the ``wrong''  behavior of the UV modes in the single stream approximation with the ``correct'' ones, encoded in the  correlators measured, for instance in N-body simulations. However, while the two-point correlator $\langle  h_a^{R,ss} \vp^{R,ss}_{b_1} \rangle'$ starts contributing to the UV loops for  $P^{R,ss}_{ab}(k)$ at 1-loop order, $\langle  h_a^{R,ss} \vp^{R,ss}_{b_1} \vp^{R,ss}_{b_2} \rangle'$ does it only from 2-loop, since it corrects the single-stream bispectrum at 1-loop order, and so on. In summary, if we want to correct the UV-loops behaviour of  the $l-$loop order PS, then we need consider only the correlators up to $\Delta \langle  h_a^{R} \vp^{R}_{b_1} \vp^{R}_{b_2} \cdots \vp^{R}_{b_l} \rangle'$, and, correspondingly, only the  first $l$ equations of the system. 

At the lowest order, we will therefore have
\beq
P^R_{ab}(k) \simeq  P^{R,1-loop}_{ab}(k) + \Delta P^{R,1-loop}_{ab}(k)  \,,
\label{P1l}
\eeq
where $P^{R,1-loop}_{ab}(k) $ is the PS computed at 1-loop SPT with the linear PS filtered at the scale $R$, while $\Delta P^{R,-1loop}_{ab}(k) $ is obtained from eq.~\re{evDP} with $\Delta B^R_{bcd}(k,p_1,p_2)=0$ and $ \Delta \langle h_a^R(\bk) \vp^R_b(-\bk) \rangle'$ subtracted at 1-loop.

In order to go to next order, one has to take into account the 2-loop PS and include eq.~\re{DB}, with $\Delta T^R=0$ and the three point correlator $\langle  h_a^{R} \vp^{R}_{b_1} \vp^{R}_{b_2} \rangle'$ subtracted at 1-loop order. 

\begin{figure}
\centering{ 
\includegraphics[width=0.49\textwidth,clip]{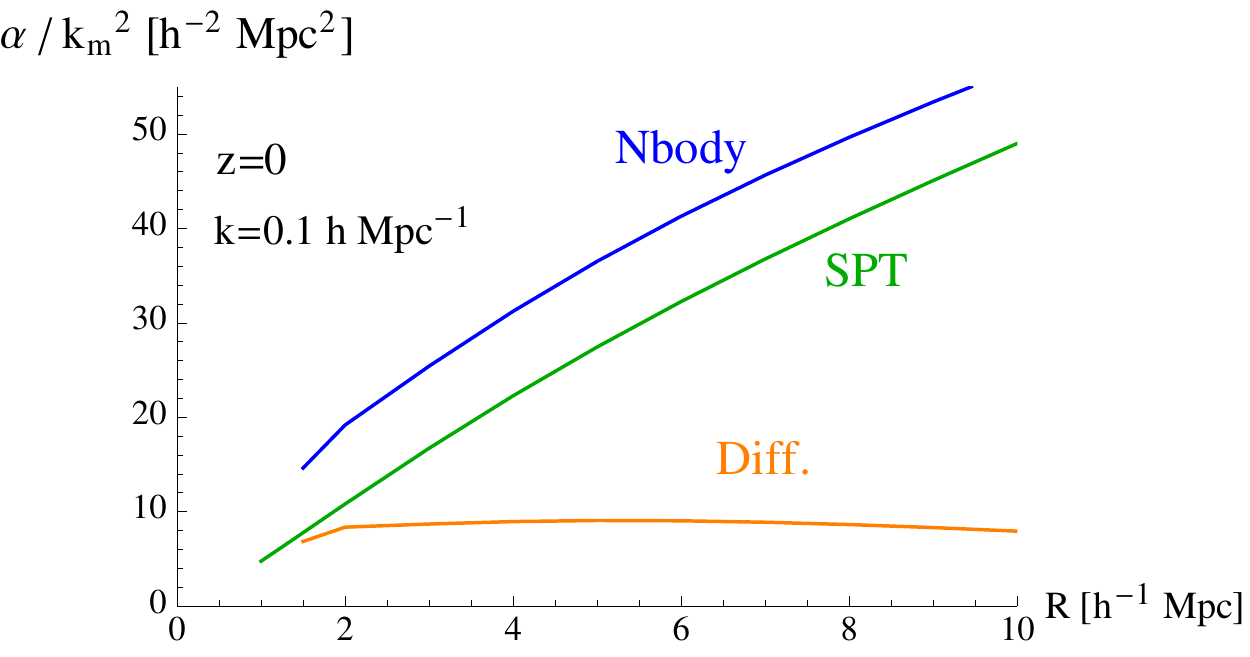}
\includegraphics[width=0.49\textwidth]{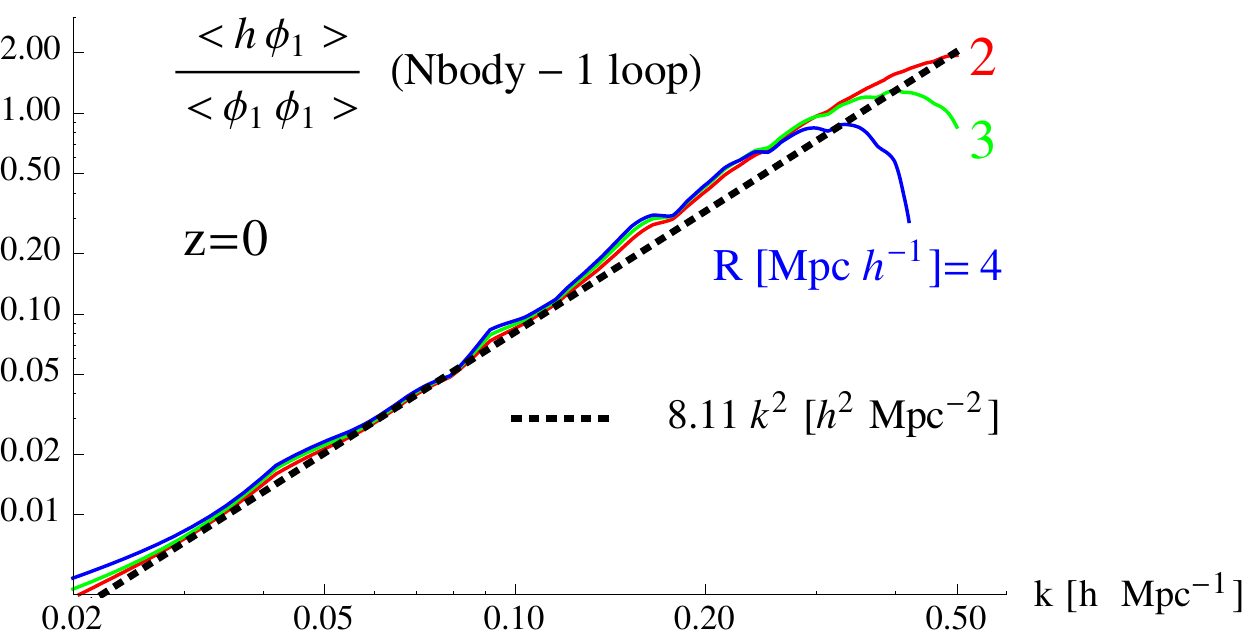}
}
\caption{Dependence on the  CG cutoff scale $R$ of the coefficient $\frac{\alpha^R}{k_m^2}$ (blue curve),  $\frac{\alpha^{R,ss}}{k_m^2}$ (green curve), and of their difference $\frac{\Delta \alpha}{k_m^2}$ (orange curve). The coefficients are obtained from $b=1$ in eq. (\ref{def-alpha}), for $k=0.1 h / {\rm Mpc}$ and at $z=0$. As discussed in the text, the difference exhibits a much smaller dependence on the cut-off scale.  Right panel: 
Scale dependence of $\frac{k^2 \, \Delta \alpha \left( z_0 \right)}{k_m^2}$ for three different values of the cut-off scale $R$. The dotted curve is the theoretical prediction $k^2$ times a $z-$dependent coefficient, obtained as explained in  \ref{app:eta-C}.
}
\label{fig:CGcutoff-scale} 
\end{figure}

As we anticipated above, the total PS, eq.~\re{P1l}, should not depend on the coarse graining scale $R$ apart from an overall dependence on the filter function. For instance, the density-density PS should depend on $R$ only through a $W[k R]^2$ factor, where $W[k R]$ is the filter function in Fourier space. The same holds for the PS computed in the single stream approximation at a finite loop order (see \cite{Manzotti:2014loa} for a 1-loop check). Therefore, one has to check that no spurious $R$-dependence is induced by the difference between the source correlator measured  in simulations and that computed in SPT. Indeed, each of them, taken separately, has a strong $R$-dependence, as shown in the left panel of fig.~\ref{fig:CGcutoff-scale}, where we use the parameterization
\beq
 \langle h_a^R(\bk) \vp^R_b(-\bk) \rangle'  = \alpha^R(\eta) \frac{k^2}{k_m^2} P^R_{1b}(k;\eta)\,\delta_{a2}\,,
 \label{def-alpha}
 \eeq
and we plot the quantities $ \alpha^R(\eta)$, $\alpha^{R,ss}(\eta)$, and the difference between the two (for $b=1$).  
The simulations we use were presented already in   \cite{Manzotti:2014loa}, see  \ref{app:eta-C} for details.  They are based on the TreePM code GADGET-II \cite{Gadget_II}. They follow the evolution, until z = 0, of $n_{part} = 512^3$ CDM particles within a periodic box of $L_{box} = 512 h^{-1} \mathrm{Mpc}$ comoving. The initial conditions were generated at z = 99 by displacing the positions of the CDM particles, that were initially set in a regular cubic grid, using the Zel'dovich approximation. We assumed the ``REF'' cosmological parameters  of  \cite{Manzotti:2014loa}, namely, $\Omega_m=0.271$, $\Omega_b=0.045$, $\Omega_\Lambda=0.729$, $h=0.703$, $n_s=0.966$, and $A_s=2.42\times 10^{-9}$.

As we see, for the scales of interest, the $R$-dependence of the correlator measured in N-body simulations is mostly cancelled by the perturbative one already at 1-loop, so that a $R$-independent  function,  
\begin{equation}
\Delta \alpha(\eta) =  \alpha^R(\eta) - \alpha^{R,ss}(\eta) \;, 
\label{delta-alpha}
\end{equation} 
can be defined.  The residual $R$-dependence, is given by two contributions: nonlinear effects from scales $q\agt 1/R$ not captured by the 1-loop subtraction, and non-perturbative (that is, beyond single-stream) effects from scales $q\alt 1/R$. The magnitude of these effects decreases with the external momentum  $k$, and, as we see from this plot, they are clearly subdominant for $k$ in the BAO range of scales. The complementary information to the left panel of Fig.~\ref{fig:CGcutoff-scale} is given in the right panel of the figure, where the scale-dependence of the ratio between the correlator and the PS is given, and the $k^2$ dependence clearly emerges.

Notice that the $\Delta \alpha(\eta)$ function bears some analogies with the sound speed coefficient in the EFToLSS \cite{Carrasco:2012cv}, however, with some key differences, that we now outline. First of all, the stress tensor in \cite{Carrasco:2012cv}, whose divergence gives basically our source terms, eq.~ \re{UVsources}, is expanded in terms of the {\em linear} filtered fields, and therefore the coefficients of this expansion are obtained from the cross-correlator of the sources with the linear fields. Here, on the other hand, the relevant cross-correlators involve the sources and the {\em nonlinear} fields $\vp_a^R$, and therefore include effects, like short scale displacements and source-source correlators ( $\langle h^R_a h^R_b \rangle'$, where the second $h^R_b$ is contained in the nonlinear evolution of the $\vp_b^R$ field) which are not included at 1-loop order in EFToLSS. Moreover, we will directly measure these sources, and the coefficient $\Delta \alpha(\eta)$, from N-body simulation and then include the result into our evolution equations for the PS. While this procedure can be followed also in the EFToLSS, more often, in practical applications, one first derives an expression for the PS containing the ``sound speed'' and other counterterms  as parameters to be fitted from the PS measured in simulations. In doing so, the physical meaning of these counterterms is less transparent, and the amount of ``overfitting'', in order to get the PS right, is difficult to estimate.

In summary, we can choose $R$ in the plateau region, or equivalently, take the formal $R\to 0$ limit, and consider the evolution equation
\beqra
&&\!\!\!\!\!\! \!\!\!\!\!\! \!\!\!\!\!\! \!\!\!\!\!\! \partial_\eta \Delta P_{ab}(k;\eta) = \Bigg[ -\Omega_{ac}\, \Delta P_{cb}(k;\eta) \nonumber\\
&&  \qquad- \Delta \alpha(\eta) \frac{k^2}{k_m^2} \left[P^{1-loop}_{1b}(k;\eta)+\Delta P_{1b}(k;\eta)\right]\,\delta_{a2} +(a\leftrightarrow b)\Bigg]\,.
\label{UVeq}
\eeqra

\begin{figure}
\centering{\includegraphics[width=0.3\textwidth,clip]{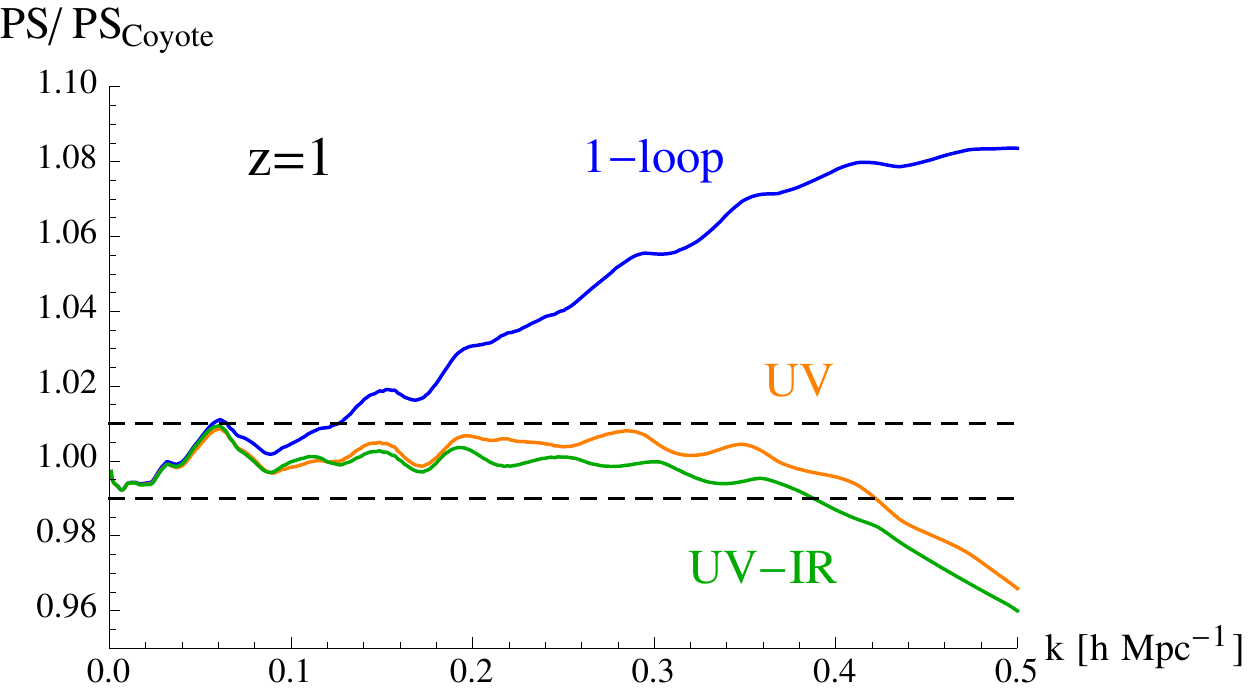}
\includegraphics[width=0.3\textwidth,clip]{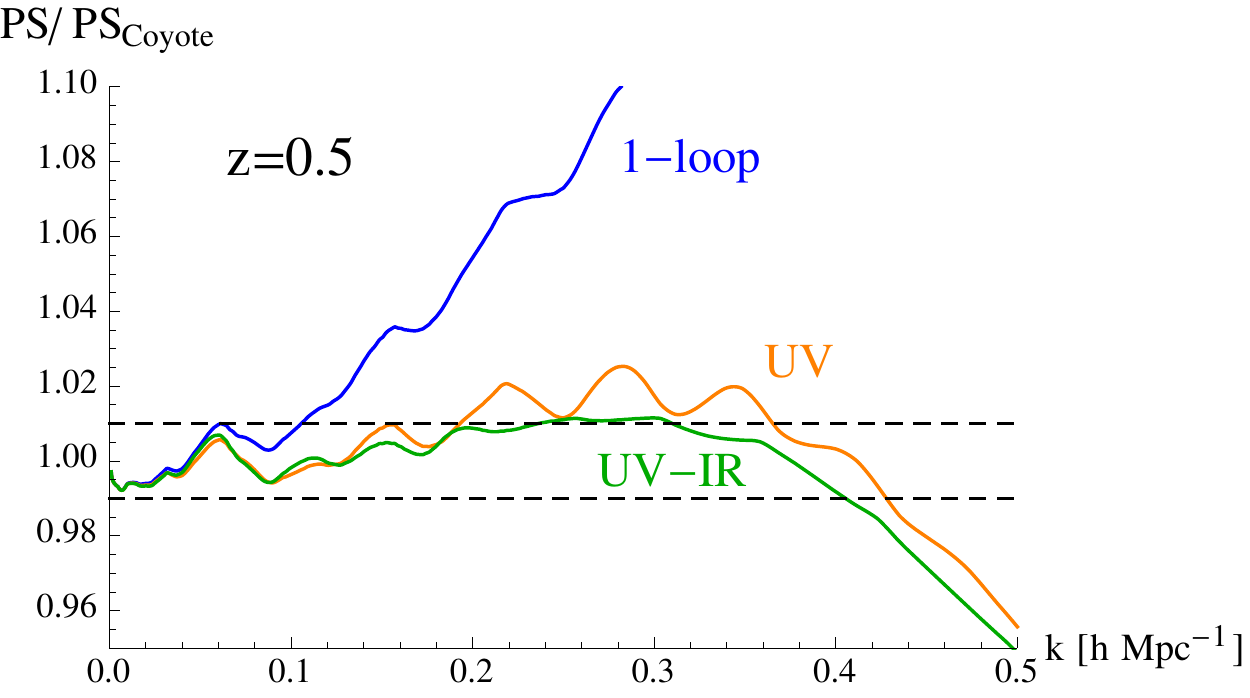}
\includegraphics[width=0.3\textwidth,clip]{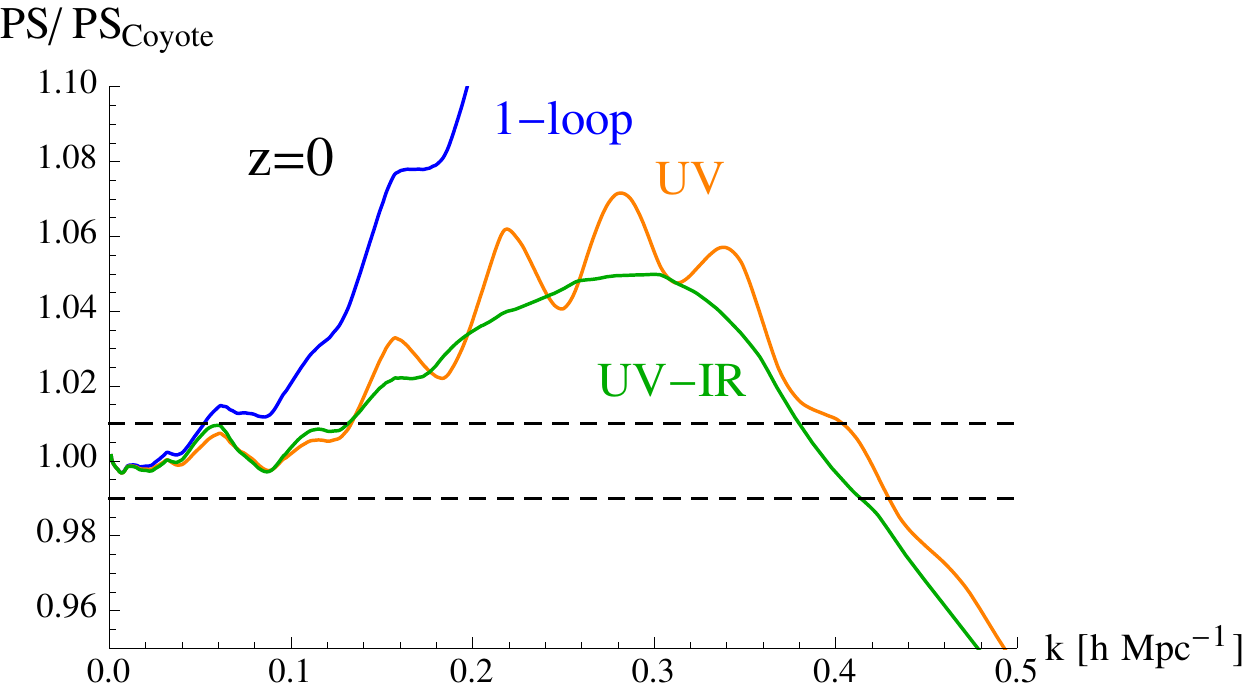}
}
\caption{PS from different computational schemes divided by the Coyote PS. The blue curve is the 1-loop SPT result. The orange curve is obtained from the UV-improved ETRG system (\ref{UVeq}). The green curve is  the UV- and IR-improved result (\ref{PS-tot}). The different panels correspond to different redshifts. The UV-improved curve performs substantially better than the SPT result, however it is still does not properly reproduces the BAO oscillations, which are instead well reproduced by the UV- and IR-improved result. The horizontal dashed lines show the band for which our results differ less than $\pm 1\%$ from the Coyote PS. 
}
\label{PStotal}
\end{figure}

In fig.~\ref{PStotal} we show the ratios between eq.~\re{P1l}  (orange line) and the PS computed with the Coyote interpolator of N-body simulations \cite{Heitmann:2013bra}. The agreement clearly improves over the 1-loop SPT result (blue line) showing that the UV correction represented by the $h_a^R$ sources plays a decisive role, already at the lowest order considered here, namely, correcting the UV of the 1-loop PS.  At $k=0.1 \, h \, {\rm Mpc}^{-1}$, the source correlator modifies the PS by $\simeq -0.6 \%$ at $z=1$, by  $\simeq -1.1 \%$ at $z=0.5$, and by $\simeq -1.8 \%$ at $z=0$. The agreement in the PS shape degrades at low redshifts, where higher loop orders should be taken into account along the lines discussed  below eq.~\re{DB}.  However, in the following, we focus our attention on the residual BAO oscillations exhibited by the orange lines, which indicate that only improving the UV effects does not account for the BAO damping well enough. In the next section we will discuss how to deal with this issue.

\section{IR resummation and BAO wiggles}
\label{IRsum}
The damping of the BAO wiggles in the PS, and of the corresponding peak in the correlation function is mainly caused by random long range displacements \cite{Eisenstein:2006nj}.  It is well known that such effects are badly reproduced in SPT at any finite order, while they are much better taken into account by the Zel'dovich approximation, which provides a resummation at all SPT orders (see for instance, \cite{Peloso:2015jua}). In our approach, the effect of these long range displacements on an intermediate scale $k$ are encoded in the last term at the RHS of eq.~\re{IRsplit} when the momentum of the velocity field is $q_1\ll k$. We will now discuss how they can be naturally resummed.

\begin{figure}
\centering{ 
\includegraphics[width=1\textwidth,clip]{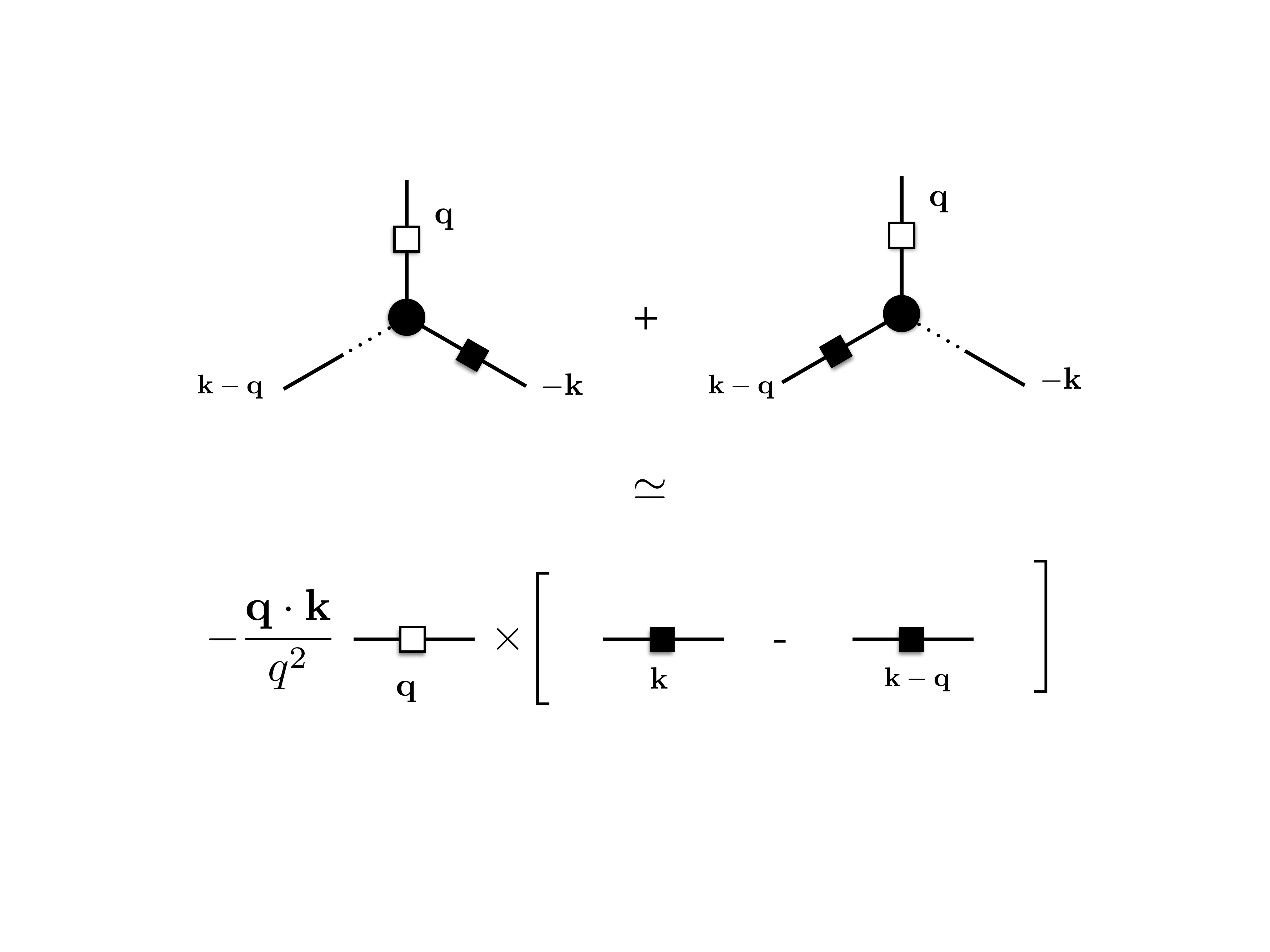}}
\caption{The consistency relations allow the bispectra appearing in eq.~\re{dispeff} to be rewritten as the product of a linear PS, $P^0 \left( q \right)$,  and a difference of nonlinear PS.}
\label{bisp3}
\end{figure}

Indeed, if one applies again the equation of motion, eq.~\re{eom}, to the correlator $\langle \vp_a(\bk,\eta) \vp_b(-\bk,\eta)\rangle^\prime$, to get the evolution equation for the PS, then, the last term at the RHS of eq.~\re{IRsplit} gives
\beqra
   &&\!\!\!\!\!\!\!\!\!\!\!\!\!\!\!\!\!\!\!\!\!\!\!\!\!\! e^\eta\; \int\frac{d^3 q}{(2 \pi)^3}\;\frac{\bk\cdot\bq}{q^2} \Bigg[ \langle \vp_2(\bq)\vp_a(\bk-\bq) \vp_b(-\bk) \rangle^\prime
   +\langle \vp_a(\bk) \vp_b(-\bk+\bq)\vp_2(-\bq)  \rangle^\prime \Bigg]\,.
   \label{dispeff}
\eeqra
The consistency relations first derived in \cite{Peloso:2013zw,Kehagias:2013yd} for the bispectrum  give
\beq
 \!\!\!\!\!\!\!\!\!\!\!\!\!\!\!\! \!\!\!\!\!\!\!\!\!\!\!\!\!\!\!\!  \langle \vp_2(\bq)\vp_a(\bk-\bq) \vp_b(-\bk) \rangle^\prime \simeq - e^\eta \frac{\bk\cdot\bq}{q^2} P^0(q) \left(P_{ab}(k)-P_{ab}(|\bk-\bq|) \right) +O\left(\left(\frac{q}{k}\right)^0\right),
 \label{newcr}
\eeq
in the $q\ll k$ configuration. The consistency relation is depicted diagrammatically in fig.~\ref{bisp3}. The second term in \re{dispeff} gives the same contribution.

Notice that, while $P^0(q)$ in eq.~\re{newcr} is taken to be the linear PS, the other two are fully nonlinear. Moreover, the RHS vanishes  at the leading order in $q/k$, that is, if one sets the argument of the second PS inside parentheses to $k$. This is in agreement with the form of the consistency relations derived in \cite{Peloso:2013zw}, which vanish if, as in this case, all the fields are taken at equal times. However, as we now show, the different arguments of the two PS is crucial when they  have an oscillatory component. 

Inserting \re{newcr} in \re{dispeff} gives
\beqra
&& -2 e^{2\eta} \int^{\Lambda(k)} \frac{d^3 q}{(2 \pi)^3} \left(\frac{\bk\cdot\bq}{q^2}\right)^2 \,P^0(q) \left(P_{ab}(k)-P_{ab}(|\bk-\bq|) \right)\nonumber\\
&&\simeq  -2 e^{2\eta} \frac{k^2}{(2\pi)^2} \int^{\Lambda(k)}  dq\,P^0(q)\int_{-1}^1 dx\,x^2 \left(P_{ab}(k) - P_{ab}(k-q x)\right)\nonumber\\
&&= -2 e^{2\eta} \frac{k^2}{(2\pi)^2} \int^{\Lambda(k)}  dq\,P^0(q) \bar P^{1}_{ab}(k;q) \, F^1(q\; r_{bao})
\label{sigma}
\eeqra
where we have inserted a UV cutoff $\Lambda(k) \alt k$,  in order to enforce the validity range of the consistency relation \re{newcr}.   We have defined
\beq
\bar P^{n}_{ab}(k;q)\equiv \frac{\int_{-1}^1 dx\,x^{2n} \left(1-\frac{P_{ab}(k-q x) }{P_{ab}(k)} \right)}{F^n(q\; r_{bao}) }P_{ab}(k),
\label{Pwd}
\eeq
for $q\alt k$,
with
\beq
F^n(q\; r_{bao}) \equiv \int_{-1}^1 dx \,x^{2n}\left(1-\cos(q\, r_{bao} \,x) \right),
\label{Fn-def}
\eeq
so that, $F^1(q\; r_{bao})=2\, ( 1-j_0(q\,r_{bao}) + 2 j_2(q \,r_{bao}))/3  $,
where the $j_n(x)$ are the spherical Bessel functions.

If the nonlinear PS has an oscillatory component, 
\beq
P_{ab}(k)=P^{nw}_{ab}(k) (1 + A_{ab}(k) \sin(k \,r_{bao})) \equiv P^{nw}_{ab}(k) +P^{w}_{ab}(k)\,, 
\label{Pnw-Pw}
\eeq
with $A_{ab}(k) $ a smooth modulating function which damps the oscillations beyond the Silk scale,
then eq.~\re{Pwd}  returns the oscillatory component itself plus a smooth contribution,
\beq
\bar P^{n}_{ab}(k;q) = P^{w}_{ab}(k) + O\left(P^{nw ''}_{ab}(k)/r_{bao}^2  \right)\,,
\label{Pwfind}
\eeq
and other terms proportional to derivatives of  $A_{ab}(k)$, which are suppressed since the oscillatory part is proportional to $\Omega_b/\Omega_m$.
Therefore, eq.~\re{sigma} gives
\beq
-2 e^{2 \eta} k^2\,\Xi(r_{bao}) P^{w}_{ab}(k)  + O\left( P^{nw ''}_{ab}  \right)\,,
\label{IRres}
\eeq
with
\beq
\Xi(r_{bao}) \equiv \frac{1}{6 \pi^2}\int^{\Lambda(k)}  dq\,P^0(q)\,( 1-j_0(q\,r_{bao}) + 2 j_2(q \,r_{bao}))\,.
\label{xi}
\eeq
The exact value of the cut-off $\Lambda(k)$ (slightly) affects the amplitude of the evolved BAO's but not their scale, as one can see by comparing the three red lines (dashed, solid,  and dotted) in Fig.~\ref{effects}, in which we plot the results for the oscillating part of the PS obtained by multiplying the integrand in \re{xi} by $\exp (-q/(c \,k)^2)$ with $c=1/2,1,\infty$, respectively, and integrating in $q$ from $0$ to $\infty$. It should be noted that the integrand of eq.~\re{sigma}, as far as the oscillatory part is concerned, is naturally cut-off at the Silk scale. Therefore, even removing the UV cutoff altogether (setting $c=\infty$), scales  $q \agt k_{Silk}\simeq 0.12 \,{\mathrm{h/Mpc}}$ do not contribute to the resummation.

If we now consider the equation for the PS derived in the previous section, and we add to it the IR resummation term, eq.~\re{IRres}, we have completed our goal: we have an evolution equation in which IR, intermediate, and UV scales are taken into account. 

We note that the last line of eq. (\ref{sigma}) has been obtained by multiplying and dividing the previous line by the function $F^1 \left( q \,r_{bao} \right)$ specified in eq. (\ref{Fn-def}). For PS of the form (\ref{Pnw-Pw}), the final line of eq. (\ref{sigma}) simplifies further into (\ref{Pwfind}), where the  scale $r_{bao}$, which is the comoving sound horizon at recombination,  appears.  For a given cosmology, we can compute the $r_{bao}$ using eq. (6) of \cite{Komatsu:2008hk}. However, notice that the BAO extraction procedure defined in the next section is quite insensitive to the input $r_{bao}$ value, see  discussion after eq.~\re{Rcont}. 

We can consider also the same equation for a ``smooth" cosmology, in which the initial PS has no BAO feature. They will not be generated by the evolution equation itself. If we subtract this equation from the one for the real cosmology, we get an evolution equation for the oscillatory component of the PS, in which the $ O\left( P^{nw ''}_{ab}/r_{bao}^2  \right)$ terms in \re{IRres} cancel out, which, splitting the oscillatory part as 
\beq P^{w}_{ab}(k;\eta)=P^{w,ss}_{ab}(k;\eta)+\Delta P^w_{ab}(k;\eta)\,,
\label{Pws}
\eeq gives the two equations 
\beqra
&&\!\!\!\!\!\! \!\!\!\!\!\! \!\!\!\!\!\! \!\!\!\!\!\!  \!\!\!\!\!\! \partial_\eta P^{w,ss}_{ab}(k;\eta) = \Bigg[ -\Omega_{ac}\,  P^{w,ss}_{cb}(k;\eta)-\Omega_{bc}\,  P^{w,ss}_{ac}(k;\eta) - 2 e^{2 \eta} k^2\,\Xi(r_{bao}) P^{w,ss}_{ab}(k) \Bigg]\,,\nonumber\\
&&\!\!\!\!\!\! \!\!\!\!\!\! \!\!\!\!\!\! \!\!\!\!\!\!  \!\!\!\!\!\! \partial_\eta \Delta P^w_{ab}(k;\eta) = \Bigg[ -\Omega_{ac}\, \Delta P^w_{cb}(k;\eta) \nonumber\\
&& \qquad - \alpha(\eta) \frac{k^2}{k_m^2} \left[P^{1-loop,w}_{1b}(k;\eta)+\Delta P^w_{1b}(k;\eta)\right]\,\delta_{a2} +(a\leftrightarrow b)\Bigg]\,,
\label{weqs}
\eeqra
with initial conditions $P^{w,ss}_{ab}(k;\etain)= P^{1-loop,w}_{ab}(k;\etain)$, $\Delta P^w_{ab}(k;\etain)=0$.
If $\etain$ is taken early enough, we can approximate $ P^{1-loop,w}_{ab}(k;\etain)\simeq P^{0,w}(k) u_a u_b$, and the first equation has the analytical solution
\beq
P^{w,ss}_{ab}(k;\eta)=P^{0,w}(k) u_a u_b\;\exp[-e^{2 \eta} k^2\,\Xi(r_{bao})]\,.
\eeq
The total PS is obtained by adding the smooth component, 
\beq P_{ab}(k;\eta)=P^{nw}_{ab}(k;\eta)+P^w_{ab}(k;\eta)\,,
\label{PS-tot}
\eeq with  
\beq P^{nw}_{ab}(k;\eta)=P^{nw,1-loop}_{ab}(k;\eta)+\Delta P^{nw}_{ab}(k;\eta)\,,
\label{Pwns}
\eeq where $\Delta P^{nw}_{ab}(k;\eta)$ solves an equation analogous to eq.~\re{UVeq},
\beqra
&&\!\!\!\!\!\! \!\!\!\!\!\! \!\!\!\!\!\! \!\!\!\!\!\!  \!\!\!\!\!\! \partial_\eta \Delta P^{nw}_{ab}(k;\eta) = \Bigg[ -\Omega_{ac}\, \Delta P^{nw}_{cb}(k;\eta) \nonumber\\
&&\qquad  - \alpha(\eta) \frac{k^2}{k_m^2} \left[P^{1-loop,nw}_{1b}(k;\eta)+\Delta P^{nw}_{1b}(k;\eta)\right]\,\delta_{a2} +(a\leftrightarrow b)\Bigg]\,.
\label{nweqs}
\eeqra

\section{Extracting the BAO}
\label{BAOext}
\begin{figure}
\centering{ 
\includegraphics[width=0.6\textwidth,clip]{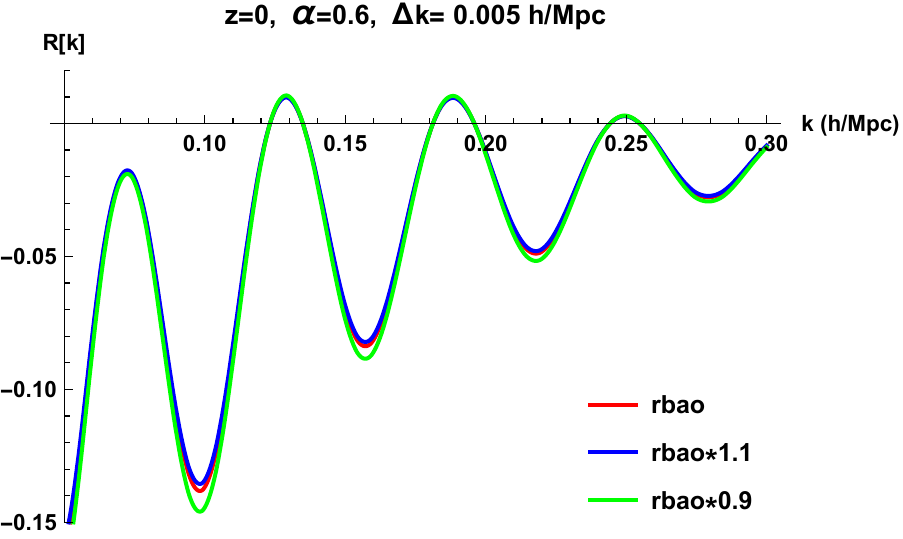}
}
\caption{The BAO extractor $R[P](k;\bar \alpha)$ applied to the linear PS using the value for $r_{bao}$ in \re{Rcont} corresponding to the one for the given cosmology (red), and to a value amplified (blue) and suppressed (green) by 10 \% with respect to it. Changing $r_{bao}$ slightly affects the amplitude of the oscillations but not the scale.
}
\label{rbao}
\end{figure}

The solution for the total PS in which the effect of the IR displacement on the oscillating component has been taken into account by eqs.~\re{weqs} are given by the green lines in fig.~\ref{PStotal} (for which we took $\Lambda = \infty$ in eq. (\ref{xi})). The residual BAO oscillations are greatly reduced with respect to the results obtained in sect.~\ref{UVsect}, while the performance on the overall PS shape (the ``broadband") is basically the same. This fact suggests that the evolutions of the two components of the PS, namely the smooth broadband shape and the oscillatory one, are governed by different physical effects: the mode-coupling with intermediate and UV scales for the former and long range displacements for the latter, with just a moderate amount of interference between them. To explore this possibility, we first look for a procedure to {\em extract} the oscillatory PS from a given PS, linear or nonlinear. Of course, this procedure has some degree of arbitrariness, as any small smooth function vanishing outside the BAO range of scales can be assigned either to the oscillatory or to the smooth part. However, the evolution equation itself suggests an optimal way to define such a procedure, as it extracts the oscillatory component which evolves (mostly) independently from the smooth one.
Indeed, as we have seen, eq.~\re{sigma} led us to define the wiggly component as in eq.~\re{Pwfind}, using eq.~\re{Pwd} (for $n=1$). Therefore we will consider the operation,
 \beqra
&& \frac{\bar P^{n}(k;q)}{P(k)} = \frac{\int_{-1}^1 dx\,x^{2n} \left(1-\frac{P(k-q x) }{P(k)} \right)}{\int_{-1}^1 dx \,x^{2n}\left(1-\cos(q\, r_{bao} \,x) \right) }\nonumber\\
&&=\frac{\int_{-\bar\alpha}^{\bar \alpha} d\alpha\; \alpha^{2n}\left(1-\frac{P(k-\frac{2 \pi}{r_{bao}}\alpha) }{P(k)}\right) }{\int_{-\bar\alpha}^{\bar \alpha} d\alpha\; \alpha^{2n}\left(1-\cos(2\pi \alpha) \right) }\equiv R[P](k;\bar\alpha,n),
\label{Rcont}
\eeqra
where the parameter $\bar\alpha$, that we will call ``range'', determines the width of the momentum interval, centered  on $k$ and expressed in units of the BAO scale, over which we take the integral. The parameter $n$, on the other hand, is set to 1 in the evolution equation, and has no effect when the operation $R[P]$ is applied on theoretical PS's, but it might be useful on real data, when different bins are measured with different errors. 
Moreover, notice that the quantity $r_{bao}$ does not affect the scale of the oscillations extracted from the PS via eq.~\re{Rcont}, as it is shown explicitly in Fig.~\ref{rbao}. In other terms, the $R[P](k;\bar\alpha,n)$ procedure extracts the scale of the true oscillations contained in  $P(k)$, that is the  $r_{bao}$ of eq.~\re{Pnw-Pw}, while the $r_{bao}$ parameter in eq.~\re{Rcont}, which could also differ from the former, only affects the amplitude of the oscillations.  

\begin{figure}
\centering{ 
\includegraphics[width=0.6\textwidth,clip]{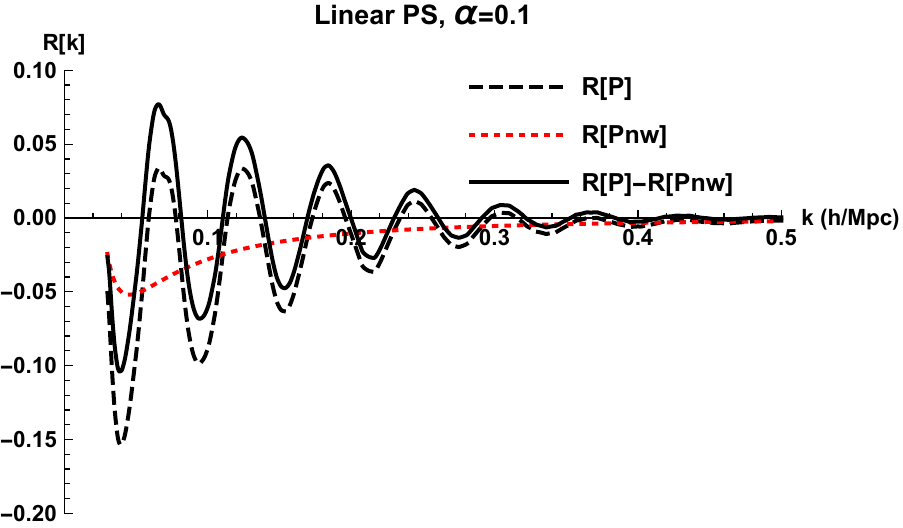}
\includegraphics[width=0.6\textwidth,clip]{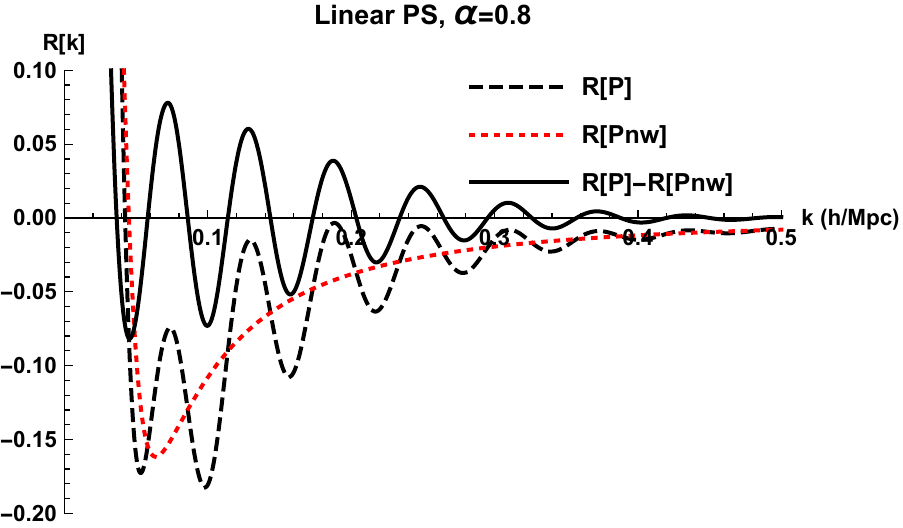}
}
\caption{The BAO extractor $R[P](k;\bar \alpha)$ applied to the linear PS (black-dashed line), to the smooth linear PS (red-dotted line), and their differences (black-solid line), for two different values of the range parameter $\bar\alpha$. 
}
\label{Rlin}
\end{figure}

As indicated in \re{Pwfind}, the ratio $ R(k;\bar\alpha,n)$ is ``contaminated'' by smooth terms coming from the non-oscillatory component of the PS, of order $\sim P''/(P r^2_{bao}) \sim 1/(k r_{bao})^2$. This is clearly seen in fig.~\ref{Rlin}, where we show the action of the operation $R[P^0]$ on the linear PS (black dashed lines), and on a smooth interpolation of it, in which BAO oscillations are absent $R[P^{0,nw}]$ (red dotted lines)\footnote{We thank Tobias Baldauf for sharing with us a code to extract the smooth PS.}. The difference between the two (black solid lines) oscillates around the x-axis.  In the following plots, we will always subtract the same $R[P^{0,nw}]$ from all the different $R[P]$, in  order to visualize oscillatory behaviors along the horizontal axis and, at the same time, keep the extraction procedure as simple as possible.

From the expression above, we can define a family of estimators for the ratio above. Assuming the data are binned, we can write
\beqra
&&\!\!\!\!\!\!\!\!\!\! \hat R[P](k_m;\bar\alpha,n) \equiv \frac{\sum_{l=-L(\bar\alpha)}^{L(\bar\alpha)} \,\left(k_{m+l} - k_m\right)^{2n} \left(1-\frac{P_{m+l}}{P_m}\right)}{\sum_{l=-L(\bar\alpha)}^{L(\bar\alpha)} \left(k_{m+l} - k_m\right)^{2n}  \left(1-\cos\left(r_{bao} \left(k_{m+l}-k_m \right)\right)\right)}\,,
\label{Rdisc}
\eeqra
where the value of the maximum $|l|$ in the sum, $L(\bar\alpha)$, is chosen such that
\beq \left| k_{m+l}-k_m \right| \le \frac{2 \pi \bar\alpha}{ r_{bao}}\;\;\;\; {\mathrm{for}}\;\; |l|\le L(\bar\alpha)\,.
\eeq
Assuming that the errors on the PS at different bins are uncorrelated, the error on $\hat R[P](k_m;\bar\alpha,n)$ is given by

\beq
\!\!\!\!\!\!\!\!\!\! \!\!\!\!\!\!\!\!\!\!\!\!\!\!\!\!\!\!\!\!  \!\!\!\! \Delta \hat R[P](k_m;\bar\alpha,n) = \frac{\sqrt{ \sum_{l\neq 0,l=-L(\bar\alpha)}^{L(\bar\alpha)} \left(k_{m+l} - k_m\right)^{4n} \left(\frac{P_{m+l}}{P_m} \right)^2\left[ \left(\frac{\Delta P_m}{P_m}\right)^2 +\left(\frac{\Delta P_{m+l}}{P_{m+l}} \right)^2 \right] }}{\sum_{l=-L(\bar\alpha)}^{L(\bar\alpha)} \left(k_{m+l} - k_m\right)^{2n}  \left(1-\cos\left(r_{bao} \left(k_{m+l}-k_m \right)\right)\right)}\,.
\label{DRdisc} 
\eeq

\begin{figure}
\centering{ 
\includegraphics[width=0.4\textwidth,clip]{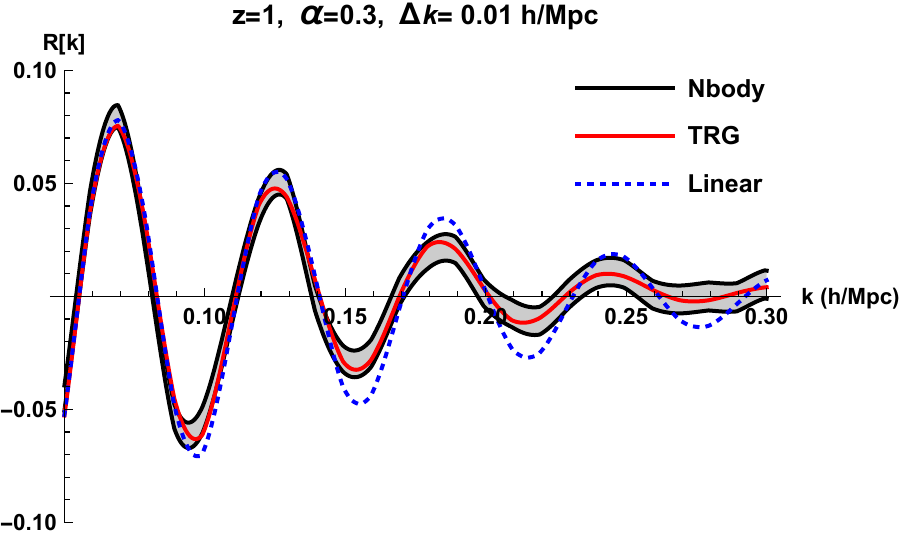}
\includegraphics[width=0.4\textwidth,clip]{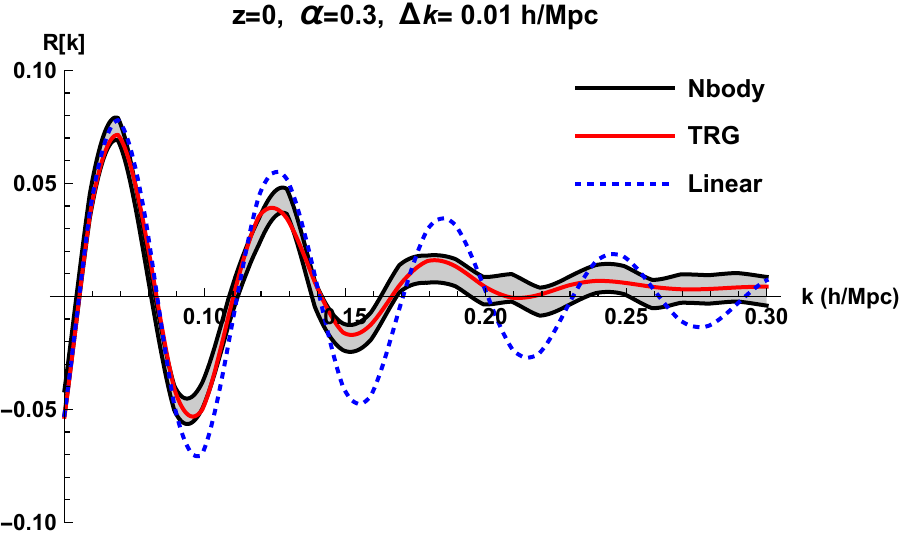}}
\centering{
\includegraphics[width=0.4\textwidth,clip]{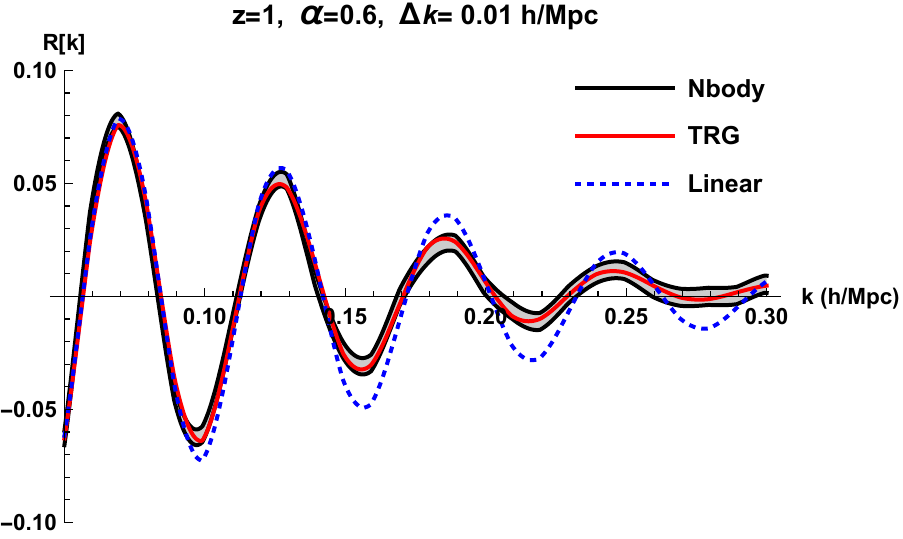}
\includegraphics[width=0.4\textwidth,clip]{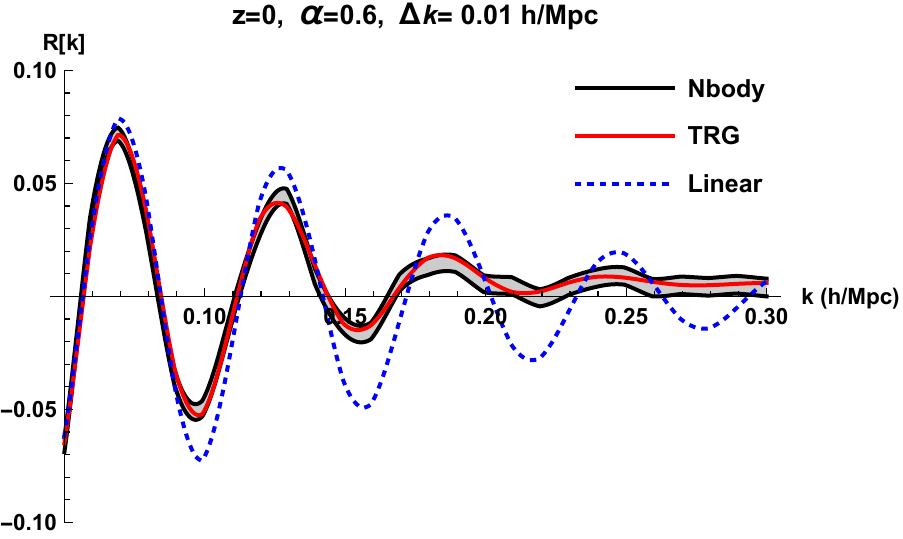}
}
\centering{ 
\includegraphics[width=0.4\textwidth,clip]{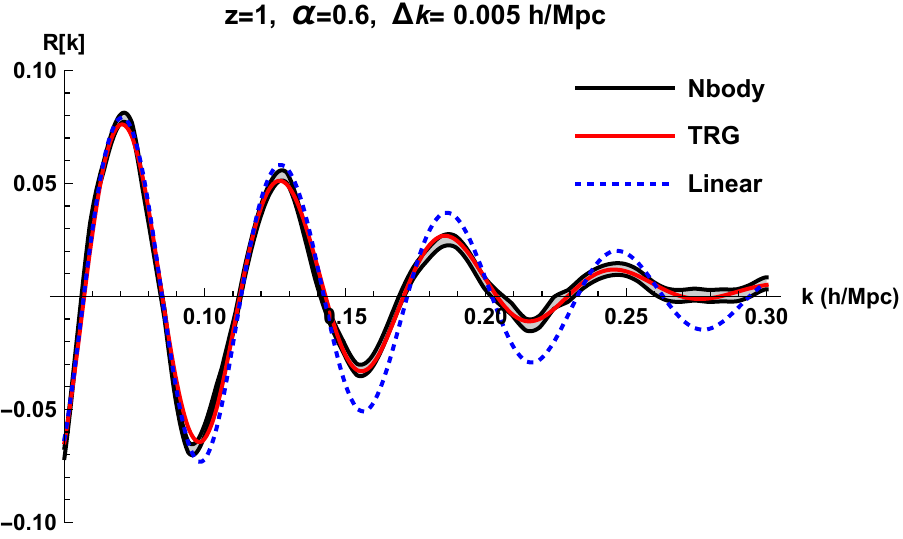}
\includegraphics[width=0.4\textwidth,clip]{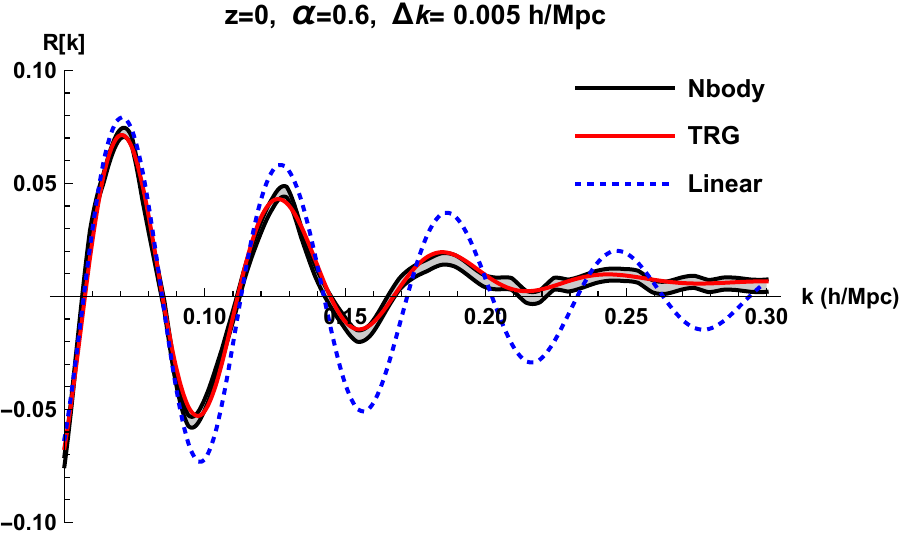}
}
\caption{The extractor $R[P](k;\bar \alpha)$ applied to the PS computed from N-body simulations (grey area) and to the TRG result described in the text at redshift $z=1$ (left column) and at $z=0$ (right column). Different values of the range parameter $\bar\alpha$ and of the binning  $\Delta k$ are shown. The grey area corresponds to assuming an error $\frac{\Delta P_m}{P_m}=1\%$  in each bin. The parameter $n$ in \re{Rcont} and \re{Rdisc} has been set to $n=0$ We also show, in blue-dashed lines, the effect of the extractor applied to the linear PS. For visualisation purposes, the same quantity $R[P^{0,nw}](k)$, where $P^{0,nw}$ is the smooth component of the linear PS, has been subtracted from all the different $R[P](k)$. }
\label{resz1}
\end{figure}

In fig.~\ref{resz1} we show the comparison between our results and the N-body simulations of ref.~\cite{Sato:2011qr}, which are optimised for the PS on the BAO scales. The UV cutoff in eq.~\re{xi} has been implemented by multiplying the integrand by  the gaussian exponential $\exp(- (q/k)^2)$. To show the behavior of the discrete $R[P]$ \re{Rdisc}, we have binned the N-body data in $\Delta k$ intervals, and we have assumed a relative error $\Delta P_m/P_m = 0.01$ for the PS in each bin (the N-body curves shown in the figure are obtained from an interpolation of the binned data 
obtained from (\ref{Rdisc}) and (\ref{DRdisc})). 

\begin{figure}
\centering{
\includegraphics[width=1\textwidth,clip]{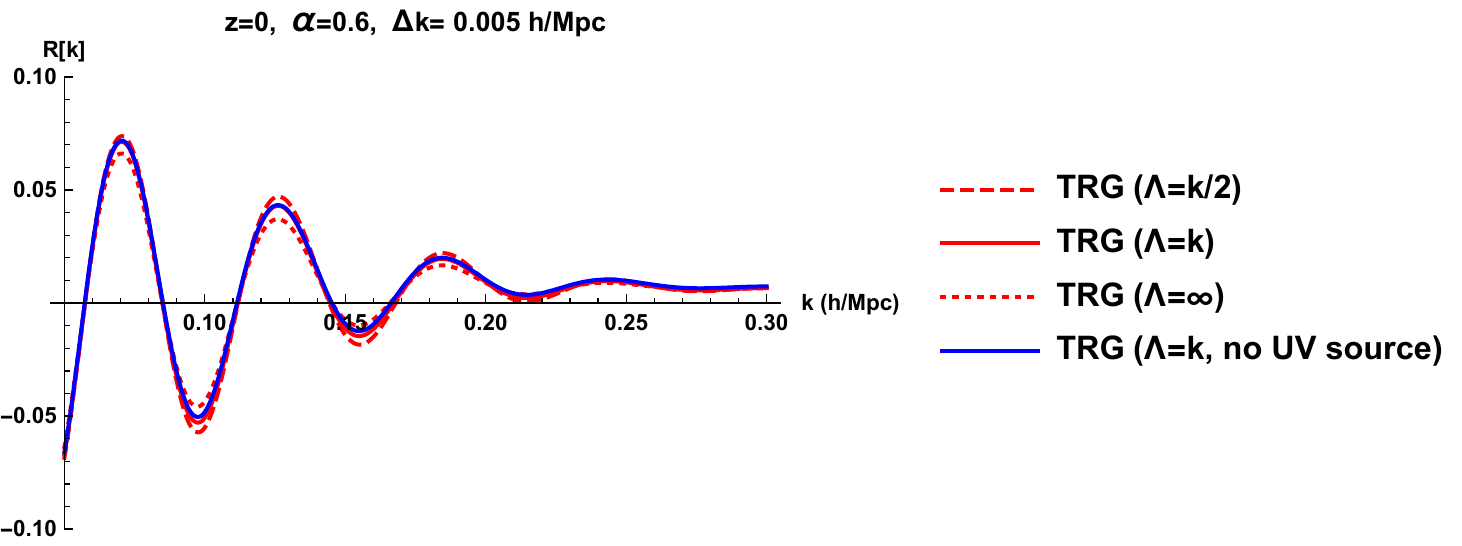}}
\caption{The effect of the UV ``sound speed'' contribution can be quantified by comparing the red and blue solid lines, which, respectively, have been obtained with and without UV sources and with the same IR contributions. The cut-off  $\Lambda(k)$ in eq.  (\ref{xi})  is implemented by multiplying the integrand by $\exp(- q^2/(c\,k)^2 )$ and by integrating in $q$ from $0$ to $\infty$. The red lines have been obtained with the same UV contribution and using different values for the $c$ constant in the IR resummation. The solid line has been obtained with $c= 1$.   The dashed and dotted lines have been obtained with, respectively,  $c=1/2$ and $c = \infty$. These results show that the BAO reconstruction method presented in this work is very robust against changes in the UV.}
\label{effects}
\end{figure}

In fig.~\ref{effects} we show the sensitivity of our results to this choice, by removing the UV cutoff entirely (that is, by taking the gaussian exponential to 1). In the same plot, we show also the effect of neglecting the UV contribution, that is, of setting $\Delta P^{w}_{ab}(k)=\Delta P^{nw}_{ab}(k)=0$ in eqs.~\re{Pws}, \re{Pwns}. This corresponds to using the 1-loop SPT results for the broadband part of the PS, which, as shown in fig.~\ref{PStotal}, reproduces the  broadband part quite badly. Nevertheless, the effect on the oscillating part is given by the difference between the red and the black lines, which is barely recognisable on this plot scale.

\section{Conclusions}
\label{conclusions}
The results of this paper  somehow pair with those of ref.~\cite{Peloso:2015jua} on the damping of the BAO peak in the correlation function. In that case, it was shown that the Zel'dovich approximation (and slights improvements of it) can account for the widening of the BAO peak which, rather than being a disturbance to be marginalised over, is a well controlled physical phenomenon which can be used to extract cosmological information. Thanks to our procedure to extract the oscillating part of the PS, one can discuss the same issue in Fourier space and show that the damping of the oscillatory component of the PS depends, in practice, only on the $\Xi$ function defined in  eq.~\re{xi}, and therefore carries information on the linear PS and the linear growth factor. These IR effects are orthogonal to the well known limitations of SPT in the UV, which are quantitatively taken into account by effective ``sound speed'', and other counterterms,  which we have shown to be basically irrelevant on the BAO evolution. 
As a result, the BAO oscillations are robustly reproduced at all scales and at all redshifts by means of the semi-analytical computation presented here, which requires only the 1-dimensional momentum integral defining the $\Xi$ function.

When extracting BAO from data one usually fits the total PS with a model containing a certain number of nuisance parameters (see, for instance, \cite{Anderson:2013zyy}). The BAO extractor introduced in \re{Rcont}, \re{Rdisc}, contains no free parameter, ($\bar \alpha$ and $n$ {\em define} an extraction procedure, they are not parameters to be fitted to compare theory and simulations or data). Moreover, since
the ratio $R$ depends on ratios of PS measured at nearby scales, it should be insensitive to most of the systematics (including bias): this will be explored in a forthcoming publication, where also the extension to redshift space will be investigated. 
The procedure can be equally applied to the reconstructed PS, obtained after the long range displacements have been undone \cite{Eisenstein:2006nk,Noh:2009bb,Tassev:2012hu,White:2015eaa}: therefore it is not an alternative to reconstruction, but rather, it provides a parameter independent procedure to extract BAO information from reconstructed data.

As for the broadband part of the PS, we showed that a 1-loop SPT computation supplemented with just one UV counterterm gives results in agreement with N-body simulations up to $k_{max}\sim0.4\;\mathrm{h/Mpc}$ for $z\ge 0.5$, rapidly degrading at lower redshifts. We have discussed how to systematically improve our approximation, by including higher order SPT corrections and more correlators between the UV sources and the density and velocity fields. The use of time-evolution equations considered here is particularly fit to deal with models beyond $\Lambda$CDM in which the boradband part of the PS carries a distinctive signature, like cosmologies with massive neutrinos or based on modified GR, as the scale-dependence of the growth factor can be directly implemented in the linear evolution matrix in eq.~\re{bigomega}.

\section*{Acknowledgments}
We thank Francisco Villaescusa-Navarro and Matteo Viel for discussions and for providing us with the N-body simulations from which the UV source terms have been extracted. We also thank M. Sato and T. Matsubara for providing us with the N-body simulations used in the comparisons in figs.~\ref{resz1}. M. Pietroni thanks T. Baldauf for discussions. E. Noda and M. Pietroni acknowledge support from the European Union Horizon 2020 research and innovation programme under the Marie Sklodowska-Curie grant agreements Invisible- sPlus RISE No. 690575, Elusives ITN No. 674896 and Invisibles ITN No. 289442.
The work of M. Peloso was supported in part by DOE grant de-sc0011842 at the University of Minnesota.

\appendix

\section{Computation of $\Delta \alpha \left( \eta \right)$} 
\label{app:eta-C}

In this Appendix we briefly discuss how we computed the coefficient $\frac{\Delta \alpha \left( \eta \right)}{k_m^2}$ 
used in the TRG equations (\ref{weqs}) and (\ref{nweqs}). We recall from equations (\ref{def-alpha}) and  (\ref{delta-alpha})  that it is obtained from the correlator between the source $h_a^R$ and the course grained fields $\vp^R$ measured from N-body simulations, minus the corresponding quantity  evaluated in single stream approximation. 

We use the N body simulations of the `reference' cosmology of   \cite{Manzotti:2014loa} (see Section 4 of that work for details). 
We use the simulations to evaluate $\frac{ \langle h_2^R \left( \bk ,\eta \right) \vp^R_1 \left( -\bk ,\eta \right) \rangle' }{
\langle \vp^R_1 \left( \bk ,\eta \right)  \vp^R_1 \left( -\bk ,\eta \right) \rangle' }$ at the redshifts $z=0,\, 0.25 ,\, 0.5 ,\, 1 ,\, 1.5 ,\, 2 ,\, 3 ,\, 5 $. 
We fix the filter scale at $R = 2 \, {\rm Mpc} \, h^{-1}$, which is the smallest value of $R$ at which the resulting $\Delta \alpha$ is in the  plateau region visible in the left panel of Figure \ref{PStotal} (we verified that using larger values of $R$ does not change the final values of $\Delta \alpha$ significantly). We then evaluate the coefficient $\frac{\alpha^R}{k_m^2}$ at $k = 0.1 \, h / {\rm Mpc}$ (as visible from the right panel of Figure  \ref{PStotal}, the result does not significantly change if we use a diffierent but comparable scale). 

From this quantity we subtract  the analogous ratio  $\frac{ \langle h_2^R \left( \bk ,\eta \right) \vp^R_1 \left( -\bk ,\eta \right) \rangle' }{\langle \vp^R_1 \left( \bk ,\eta \right)  \vp^R_1 \left( -\bk ,\eta \right) \rangle' }$, with the numerator evaluated in single stream approximation through a one loop computation, and with the Coyote Emulator \cite{Heitmann:2013bra}  PS in the denominator. The one loop computation is performed following Appendix A of  \cite{Manzotti:2014loa}.~\footnote{Specifically, we add up ``22'' and ``13'' contributions. The ``22'' contributions are obtained from the terms ${\cal C}^{h0}$ (only half of it) and ${\cal C}^{hh}$ written in eqs. (A.13) and (A.14) of   \cite{Manzotti:2014loa}; the ``13'' conttibutions are obtained from eqs. (A.20)-(A.21)-(A.22)-(A.23) of   \cite{Manzotti:2014loa}. The correlators written in  \cite{Manzotti:2014loa} are given in terms of the quantity denoted as $\vpb^{(h)}$ in that work. To obtain the correlators in terms of $h_a$ we formally performed the replacement $g_{ac} \left( \eta - s \right) \rightarrow \delta_{ac} \delta^{(D)} \left( \eta - s \right)$ in the expressions of    \cite{Manzotti:2014loa}.} 

\begin{figure}
\centering{ 
\includegraphics[width=0.49\textwidth,clip]{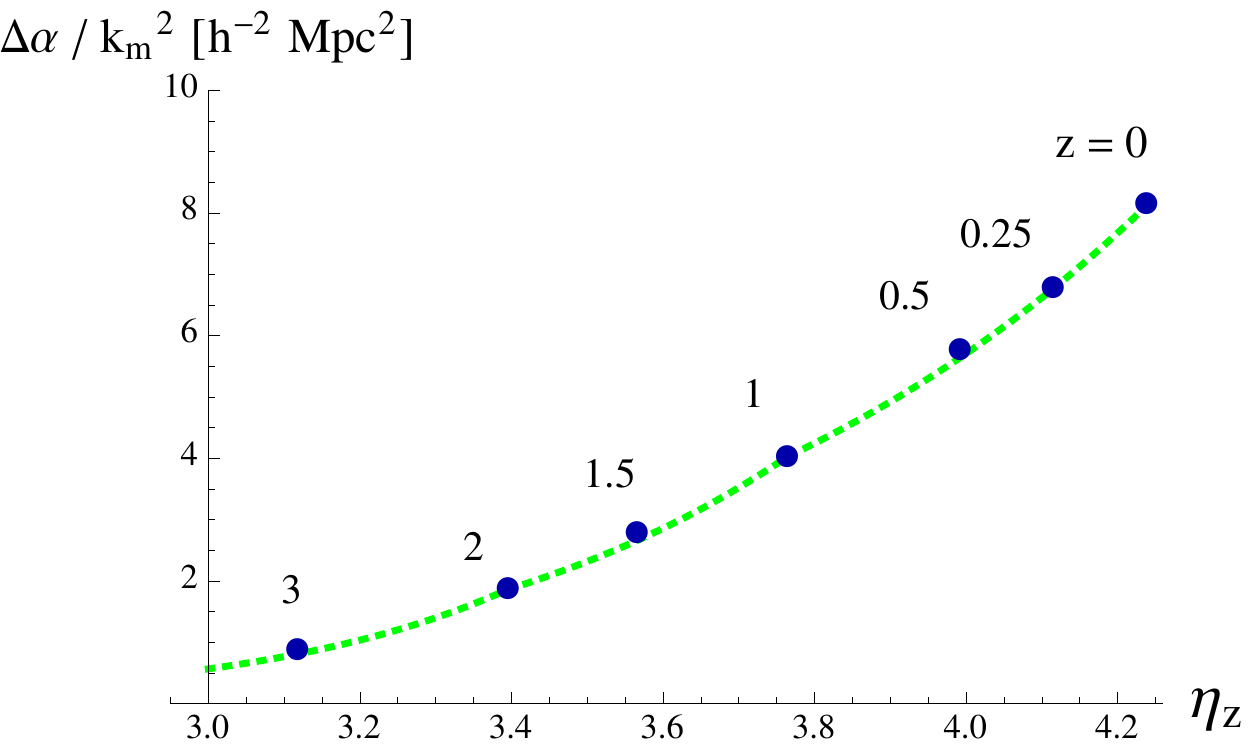}
}
\caption{Coefficient $\frac{\Delta \alpha}{k_m^2}$ as a function of the logarithm $\eta$ of the linear growth factor. The points are the data 
obtained at various redshift, as explained in the text. The numerical value next to each point indicates the corresponding redshift. The green dashed curve is the fit (\ref{C-eta}), with the coefficients ${\cal C}_0 = 8.11 \, h^{-2} \, {\rm Mpc}^2$,   $n_1 = 1.48$, and $n_2 =  2.09$. 
}
\label{fig:eta-Dalpha} 
\end{figure}

From this subtraction, we obtain the coefficient $\frac{\Delta \alpha}{k_m^2}$ at the various redshifts listed above. In the ETRG system 
a continuous function $\alpha \left( \eta \right)$ is needed.  We find that the time dependence is well reproduced by the three-parameter fit: 
\begin{equation}
\frac{\Delta \alpha \left( \eta \right)}{k_m^2} = \left\{ 
\begin{array}{l}
{\cal C}_0  \, {\rm e}^{n_1 \left( \eta - \eta_0 \right)} 
\quad\quad\quad\quad\quad\quad\quad\quad , \;\;\; \eta_1 \leq \eta \leq \eta_0 \;\; ,  \\ 
{\cal C}_0 \, {\rm e}^{n_1 \left( \eta_1 - \eta_0 \right)}  \, {\rm e}^{n_2 \left( \eta - \eta_1 \right)} 
\quad\quad\quad\quad , \;\;\;  \eta_2 \leq \eta \leq \eta_1  \;\; ,  \\ 
{\cal C}_0 \, {\rm e}^{n_1 \left( \eta_1 - \eta_0 \right)}  \, {\rm e}^{n_2 \left( \eta_2 - \eta_1 \right)} 
\, {\rm e}^{3 \left( \eta - \eta_2 \right)} 
\;\; , \;\;\;    \eta \leq \eta_2  \;\; ,   
\end{array} \right. 
\label{C-eta} 
\end{equation} 
where $\eta_i = \eta \left( z_i \right)$. This parametrization ensures that $\Delta \alpha \left( \eta \right)$ is continuous, and that it grows faster than the 1-loop single stream term at $z> z_2$ (as suggested by the fact that this subtracted quantity contains terms that are of higher order in perturbation theory. This behavior is in qualitative agreement with our  highest redshift data). We perform a fit of the data at $z=0,\, 0.25 ,\, 0.5 ,\, 1$ in a $\left\{ \log \eta ,\, \log \Delta \alpha \right\}$ plane, and obtain the best fit values ${\cal C}_0 = 8.11 \, h^{-2} \, {\rm Mpc}^2$ and $n_1 = 1.48$. We then fit the data at $z=1 ,\, 1.5 ,\, 2$ to obtain the best fit value $n_2 = 2.09$ (see Figure \ref{fig:eta-Dalpha}). The parametrization (\ref{C-eta}), with these values for the three fitting parameters ${\cal C}_0 ,\, n_1 ,\, n_2$, is used in the TRG evolutions of the main text.

\section*{References}
\bibliographystyle{JHEP}
\bibliography{/Users/massimo/Dropbox/Mnu/PostTRG/draft/mybib.bib}
\end{document}